\documentclass{article}
\usepackage{amssymb}
\usepackage{graphicx}
\usepackage{epsfig}

\newcommand{\beq}{\begin{equation}}
\newcommand{\eeq}{\end{equation}}
\newcommand{\beqn}{\begin{eqnarray}}
\newcommand{\eeqn}{\end{eqnarray}}

\begin{document}

\title{
\vspace{-2.0cm}
       {\normalsize \hfill ITEP-LAT/2002--31}   \\[-0.2cm]
       {\normalsize \hfill KANAZAWA 02--40}   \\[0.8cm]
{\normalsize\bf HEAVY QUARK POTENTIAL IN LATTICE QCD \\
AT FINITE TEMPERATURE}
\footnote{\uppercase{T}alk given by \uppercase{V}. \uppercase{B}ornyakov
at ``\uppercase{C}onfinement \uppercase{V}'', \uppercase{G}argano,
\uppercase{I}taly, 10-14 \uppercase{S}ep. 2002.}
$^,$\footnote{\uppercase{T}his work is partially supported by grants
\uppercase{INTAS}-00-00111, \uppercase{RFBR} 02-02-17308, 01-02-17456,
00-15-96-786 and \uppercase{CRDF RPI}-2364-\uppercase{MO}-02.  
\uppercase{M.C}h. is supported by \uppercase{JSPS F}ellowship \uppercase{N}o. 
\uppercase{P}01023. 
\uppercase{P.U.} is supported by \uppercase{K}anazawa foundation.}}

\author{\normalsize V. BORNYAKOV$\,{}^{\lowercase{a,b,c}}$, M. CHERNODUB$\,{}^{\lowercase{a,b}}$, Y. KOMA$\,{}^{\lowercase{a}}$, Y.~MORI$\,{}^{\lowercase{a}}$, \\
\normalsize Y.~NAKAMURA$\,{}^{\lowercase{a}}$, M. POLIKARPOV$\,{}^{\lowercase{b}}$, 
G. SCHIERHOLZ$\,{}^{\lowercase{d}}$, D. SIGAEV$\,{}^{\lowercase{b}}$, \\
\normalsize A.~SLAVNOV$\,{}^{\lowercase{e}}$, H. ST\"UBEN$\,\,{}^{\lowercase{f}}$, 
\normalsize T. SUZUKI$\,\,{}^{\lowercase{a}}$,
P. UVAROV$\,\,{}^{\lowercase{b,e}}$, \\ 
\normalsize A.~VESELOV$\,\,{}^{\lowercase{b}}$}

\date{\normalsize{\it{
${}^a$ ITP, Kanazawa University, Kanazawa, 920-1192, Japan \\
${}^b$ ITEP, B. Cheremushkinskaya 25, Moscow, 117259, Russia \\
${}^c$ IHEP, Protvino, 124280, Russia  \\
${}^d$ NIC/DESY Zeuthen, Platanenallee 6, 15738 Zeuthen, Germany and
Deutsches Elektronen-Synchrotron DESY, D-22603 Hamburg, Germany \\
${}^e$ Steklov Mathematical Institute, Vavilova 42, 117333 Moscow,
Russia \\
${}^f$ ZIB, D-14195 Berlin, Germany
}}}

\maketitle

\begin{abstract}
\noindent
Results of the study of lattice QCD with two flavors of nonperturbatively
improved Wilson fermions at finite temperature are presented.
The transition temperature for $\frac{m_{\pi}}{m_{\rho}} \sim 0.8$
and lattice spacing $a \sim 0.12$~fm is determined. A two-exponent ansatz
is successfully applied to describe the heavy quark potential
in the confinement phase.
\end{abstract}
\vspace{1cm}

Studies of $N_f=2$ lattice QCD at finite temperature with
improved actions have provided consistent estimates of 
$T_{c}$ \cite{Karsch:2000kv,AliKhan:2000iz}.
Still there are many sources of
systematic uncertainties and new computations of $T_c$  with
different actions are useful as an additional check. To make such check
we  performed first large scale simulations of the nonperturbatively
$O(a)$ improved Wilson fermion action at finite temperature.
Other goals of our work were to study the heavy quark potential
and the vacuum structure of the full QCD at $T>0$.

We employ Wilson gauge field action and fermionic action
of the same form as used by UKQCD and QCDSF
collaborations \cite{Booth:2001qp} in $T=0$ studies.
To fix the physical scale and
$\frac{m_{\pi}}{m_{\rho}}$ ratio we use their results.
Our simulations were performed on $16^3 8$ lattices for two values of
the lattice gauge coupling $\beta=5.2,\, 5.25$.

As numerical results show \cite{Karsch:2000kv} both
Polyakov loop and chiral condensate susceptibilities can be used to locate
the transition point. We use only Polyakov loop susceptibility.
We found critical temperature $T_c=213(10)$ and $222(10)$MeV
at $m_{\pi}/m_{\rho}=0.78, 0.82$, respectively.
These values are in good agreement with previous results \cite{Karsch:2000kv}
at comparable $m_{\pi}/m_{\rho}$.

To test finite size effects
simulations on $24^3\cdot 8$ lattice for $T/T_c=0.94$ have been made.  We found
that results for all our observables agree with our smaller volume results
within error bars. Thus finite size effects do not introduce strong
systematic uncertainties in our results.

The heavy quark potential $V(r,T)$ in full QCD at non-zero temperature
has been studied  in \cite{Karsch:2000kv}.
It is given by 
$
\langle L_{\vec{x}} L^{\dagger}_{\vec{y}}\rangle \slash 9 =\!\!
e^{-V(r,T)/T}
$, where $L_{\vec{x}}$ is Polyakov loop.
In the limit
$|\vec{x}-\vec{y}| \rightarrow \infty$,
$\langle L_{\vec{x}} L^{\dagger}_{\vec{y}}\rangle$ approaches the
cluster value $|\langle L\rangle|^2$,
where $|\langle L\rangle|^2 \neq 0$ because the global $Z_3$ symmetry
is broken by the fermions.

The spectral representation for the Polyakov loop correlator 
is \cite{Luscher:2002qv}
$$
\langle L_{\vec{x}} L^{\dagger}_{\vec{y}}\rangle  = \sum_{n=0}^{\infty}
w_n e^{-E_n(r)/T}.
$$
At $T=0$ one gets $V(r,T=0)=E_0(r)$. In contrast, $V(r,T)$ at $T>0$
gets contributions from all possible states.
We assume that in the confinement phase, at temperatures  below $T_c$,
the Polyakov loop correlator can be described with the help
of two states, namely string state and broken string (two static-light meson)
state:
\begin{eqnarray}
& & \hskip -6mm \frac{1}{9}\langle L_{\vec{x}} L^{\dagger}_{\vec{y}} \rangle =
 e^{-(V_0+V_{str}(r,T))/T} + e^{-2E(T) /T}\,,
\label{two_exp}
\\
& & \hskip -6mm
V_{str}(r,T) = {1\over 6r} {\rm arctan} x -\frac{\pi}{12r}
+ \sigma(T)r+{ xT\over 3} {\mathrm {arctan}}{1\over x} +{T\over 2} \ln \Bigl(1+x^2 \Bigr),
\label{pot_string}
 \\ 
 & & \hskip -6mm 
E(T)=V_0/2 + m(T)\,,
\label{meff} 
\end{eqnarray}

\noindent
where $m(T)$ is the effective quark mass at finite temperature, $x=2rT$. The
$T\neq0$ string potential (\ref{pot_string}) was derived
in \cite{Gao:kg}.
The alternative fit of our data can be done using the finite temperature
QCD static potential \cite{Karsch:1987pv}:
\beq
V_{KMS}(r,T)=\frac{\tilde{\sigma}}{\mu}(1-e^{-\mu r}) - \frac{\alpha}{r} e^{-\mu r}\,,
\label{kms}
\eeq
where $\tilde{\sigma}$, $\mu$ and $\alpha$ are parameters.
We used function (\ref{kms}) to fit the data.

\begin{figure}[tbh]
\hbox{
\hspace{-0.4cm} \epsfxsize=4.2cm \epsfysize=3.6cm \epsfbox{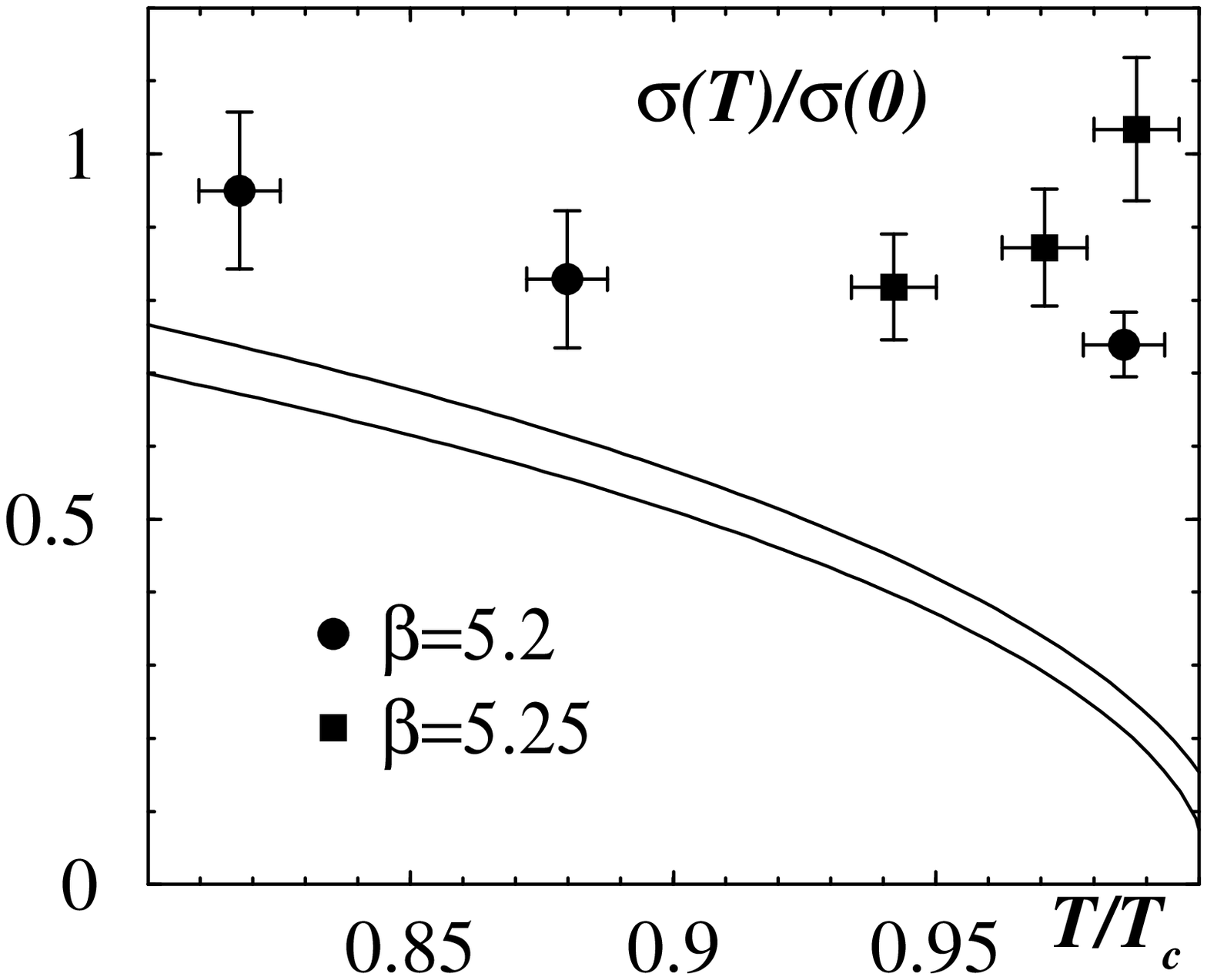}
\hspace{-0.3cm} \epsfxsize=4.0cm \epsfysize=3.6cm \epsfbox{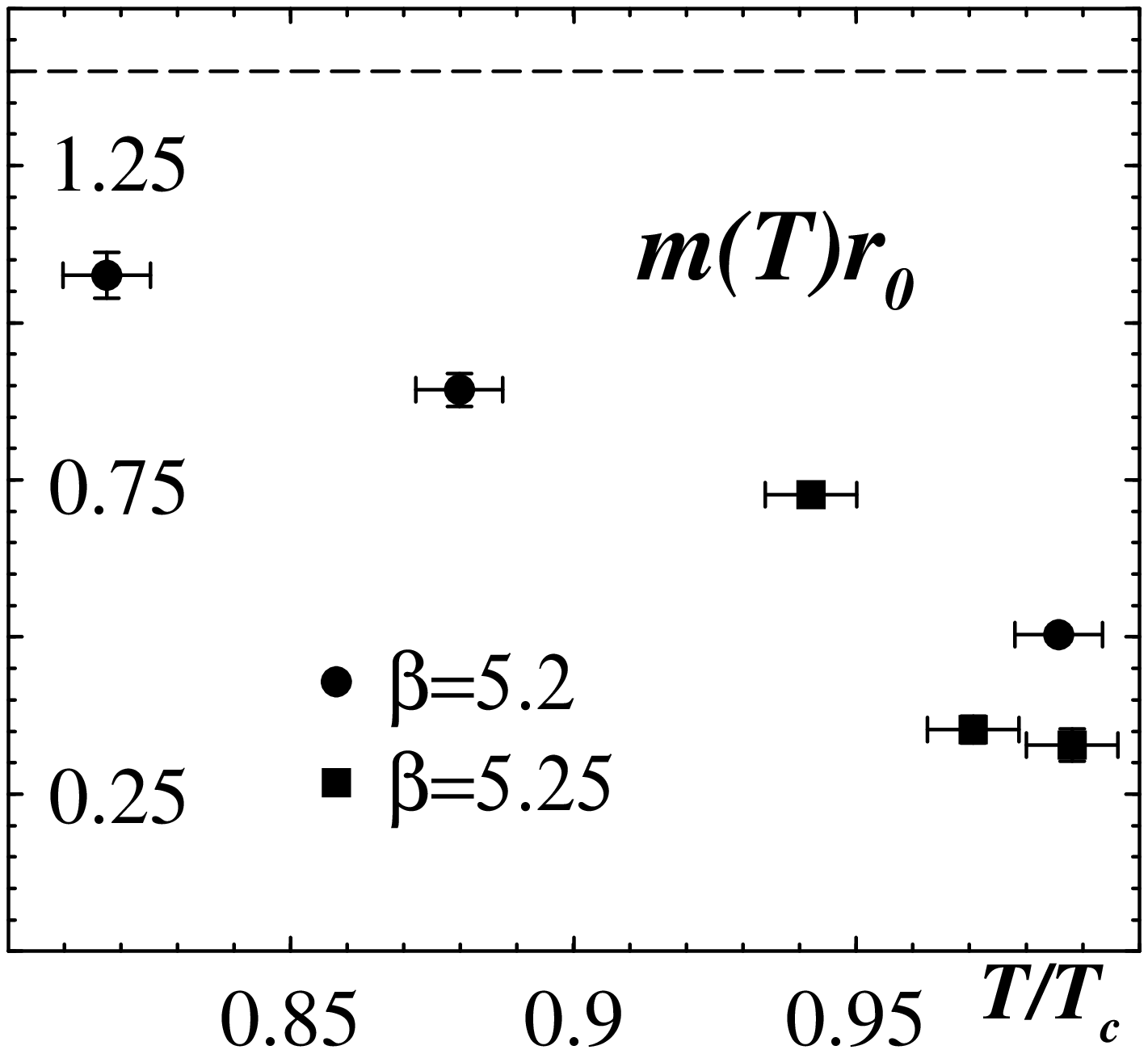}
\hspace{-0.3cm} \epsfxsize=4.2cm \epsfysize=3.6cm \epsfbox{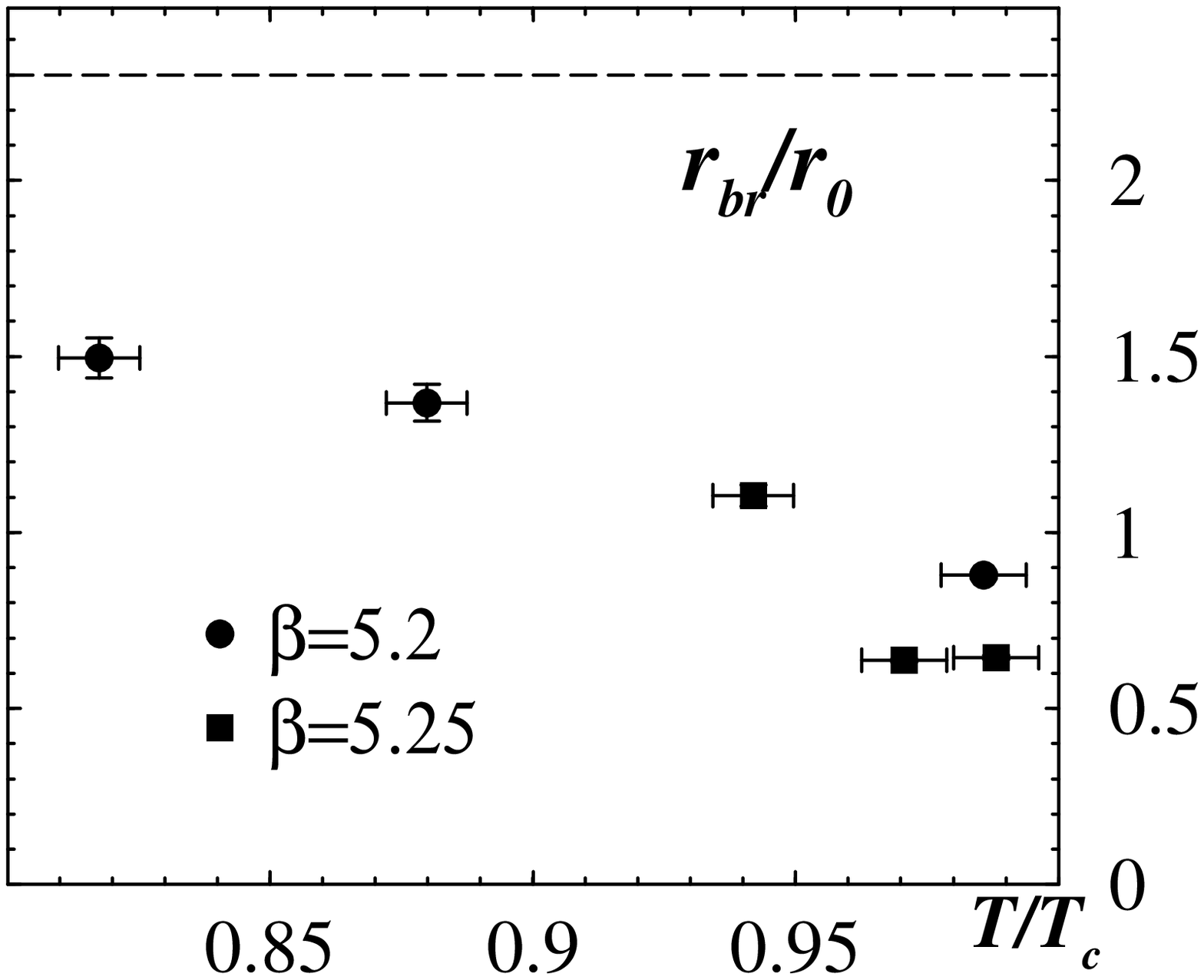}}
\caption{Best fit parameters for fit eq.(\ref{two_exp})
as functions of temperature. Solid line on the left-hand figure show quenched
results$^9$. 
Dashed horizontal lines show $T=0$ results. $r_0=0.5$ fm.} 
\label{sigma}
\vskip -2mm
\end{figure}
In computation of the Polyakov loops correlator to reduce statistical 
errors hypercubic blocking \cite{Hasenfratz:2001hp} has been employed.
Details of this computation were reported in \cite{Bornyakov:2002iv}.
Parameters of the fit (\ref{two_exp})-(\ref{meff}) are presented in 
Fig.~\ref{sigma}.
The values for the ratio $\sigma(T)/\sigma(0)$
are higher than those obtained in quenched 
QCD \cite{Kaczmarek:2000mm}, especially close to $T_c$.
The values for $m(T)$ are also 20-30 \% higher than
those obtained in \cite{Digal:2001iu}.
Using parameters of the potential
we calculate the string breaking distance $r_{sb}$
from  relation
$V_{str}(r_{sb},T) = 2m(T)$.
{}In Fig. \ref{sigma} one can see that $r_{sb}$ decreases down to values
$\sim 0.3$ fm  when temperature approaches critical value. Our fit using
$V_{str}(r,T)$, eq.(\ref{pot_string}), is probably not valid when $r_{sb}$ 
becomes so small. It still
provides reasonable values for string tension and effective mass for 
$T/T_c < 0.95$ when $r_{sb} > 0.5 fm$.
The comparison of two fits,  
eqs.(\ref{two_exp}--\ref{meff}) and
eq.(\ref{kms}) showed that both fits are equally good within our error
bars. There is an indication that with more precise data one can discriminate 
between these two fits at low temperatures. We found also that parameters
of the fit eq.(\ref{kms}) are in a clear disagreement with parameters suggested
in \cite{Wong:2001uu}.

\end{document}